\documentclass{elsart}
\usepackage{epsfig}
\usepackage{amssymb, amsmath}

\newcommand{\med}{Mediterranean Sea}
\newcommand{\pmt}{photomultiplier tube}

\begin{document}

\title{Study of Large Hemispherical Photomultiplier Tubes for the ANTARES Neutrino Telescope}

\renewcommand{\thefootnote}{\fnsymbol{footnote}}
\begin{center}          
{\bf The ANTARES Collaboration}\\
\vspace{1mm}
\begin{small}
\author[ific]{J.A.~Aguilar},
\author[grphe]{A.~Albert},
\author[dip-roma]{F.~Ameli},
\author[lam]{P.~Amram},
\author[dip-genova]{M.~Anghinolfi},
\author[erlan]{G.~Anton},
\author[dapnia]{S.~Anvar},
\author[dapnia]{F.E.~Ardellier-Desages},
\author[cppm]{E.~Aslanides},
\author[cppm]{J-J.~Aubert},
\author[oxford]{D.~Bailey},
\author[cppm]{S.~Basa},
\author[dip-genova]{M.~Battaglieri},
\author[dip-bologna]{Y.~Becherini},
\author[dip-bari]{R.~Bellotti},
\author[dapnia]{J.~Beltramelli},
\author[cppm]{V.~Bertin},
\author[cppm]{M.~Billault},
\author[grphe]{R.~Blaes},
\author[com]{F.~Blanc},
\author[dapnia]{N.~de Botton},
\author[lam]{J.~Boulesteix},
\author[nikhef]{M.C.~Bouwhuis},
\author[oxford]{C.B.~Brooks},
\author[leeds]{S.M.~Bradbury},
\author[nikhef]{R.~Bruijn},
\author[cppm]{J.~Brunner},
\author[dip-catania]{G.F.~Burgio},
\author[dip-bari]{F.~Cafagna},
\author[cppm]{A.~Calzas},
\author[dip-roma]{A.~Capone},
\author[dip-catania]{L.~Caponetto},
\author[ific]{E.~Carmona},
\author[cppm]{J.~Carr},
\author[sheffield]{S.L.~Cartwright},
\author[pisa]{E.~Castorina},
\author[pisa]{V.~Cavasinni},
\author[dip-bologna,tesre]{S.~Cecchini},
\author[llv]{P.~Charvis},
\author[dip-bari]{M.~Circella},
\author[nikhef]{C.~Colnard},
\author[brest]{C.~Comp\`ere},
\author[infn-catania-lns]{R.~Coniglione},
\author[oxford]{S.~Cooper},
\author[cppm]{P.~Coyle},
\author[dip-genova]{S.~Cuneo},
\author[brest]{G.~Damy},
\author[nikhef]{R.~van~Dantzig},
\author[llv]{A.~Deschamps},
\author[dip-bari]{C.~De~Marzo},
\author[dapnia]{D.~Denans},
\author[cppm]{J-J.~Destelle},
\author[dip-genova]{R.~De~Vita},
\author[cppm]{B.~Dinkelspiler},
\author[infn-catania-lns]{C.~Distefano},
\author[toulon]{J-F.~Drogou},
\author[dapnia]{F.~Druillole},
\author[nikhef]{J.~Engelen},
\author[grphe]{J-P.~Ernenwein},
\author[pisa]{E.~Falchini},
\author[cppm]{S.~Favard},
\author[cppm]{F.~Feinstein}\footnote{Now at: Groupe d'Astroparticules de Montpellier, UMR 5139-UM2/IN2P3-CNRS,  Universit\'e Montpellier II, Place Eug\`ene Bataillon - CC85, 34095 Montpellier  Cedex 5, France},
\author[ires]{S.~Ferry},
\author[brest]{D.~Festy},
\author[pisa]{V.~Flaminio},
\author[oxford]{J.~Fopma},
\author[com]{J-L.~Fuda},
\author[ires]{J-M.~Gallone},
\author[dip-bologna]{G.~Giacomelli},
\author[grphe]{N.~Girard},
\author[dapnia]{P.~Goret},
\author[erlan]{K.~Graf},
\author[cppm]{G.~Hallewell},
\author[erlan]{B.~Hartmann},
\author[nikhef]{A.~Heijboer},
\author[llv]{Y.~Hello},
\author[ific]{J.J.~Hern\'andez-Rey},
\author[toulon]{G.~Herrouin},
\author[erlan]{J.~H\"o{\ss}l},
\author[ires]{C.~Hoffmann},
\author[dapnia]{J.~R.~Hubbard},
\author[cppm]{M.~Jaquet},
\author[nikhef]{M.~de~Jong},
\author[dapnia]{F.~Jouvenot},
\author[erlan]{A.~Kappes},
\author[erlan]{T.~Karg},
\author[cppm]{S.~Karkar},
\author[dapnia]{M.~Karolak},
\author[erlan]{U.~Katz},
\author[cppm]{P.~Keller},
\author[nikhef]{P.~Kooijman},
\author[sheffield]{E.V.~Korolkova},
\author[apc,dapnia]{A.~Kouchner},
\author[erlan]{W.~Kretschmer},
\author[erlan]{S.~Kuch},
\author[sheffield]{V.A.~Kudryavtsev},
\author[dapnia]{H.~Lafoux},
\author[cppm]{P.~Lagier},
\author[erlan]{R.~Lahmann},
\author[dapnia]{P.~Lamare},
\author[dapnia]{J-C.~Languillat},
\author[erlan]{H.~Laschinsky},
\author[com]{L.~Laubier},
\author[cppm]{T.~Legou},
\author[brest]{Y.~Le Guen},
\author[dapnia]{H.~Le~Provost},
\author[cppm]{A.~Le~Van~Suu},
\author[dip-catania]{L.~Lo~Nigro},
\author[dip-catania]{D.~Lo~Presti},
\author[dapnia]{S.~Loucatos},
\author[dapnia]{F.~Louis},
\author[itep]{V.~Lyashuk},
\author[lam]{M.~Marcelin},
\author[dip-bologna]{A.~Margiotta},
\author[llv]{C.~Maron},
\author[toulon]{A.~Massol},
\author[dip-roma]{R.~Masullo},
\author[brest]{F.~Maz\'eas},
\author[lam]{A.~Mazure},
\author[sheffield]{J.E.~McMillan},
\author[infn-catania-lns]{E.~Migneco},
\author[com]{C.~Millot},
\author[leeds]{A.~Milovanovic},
\author[cppm]{F.~Montanet},
\author[dip-bari]{T.~Montaruli}\footnote{Now at: University of Wisconsin, Dep. of Physics, Chamberlin Hall, Madison, WI 53706, USA},
\author[brest]{J-P.~Morel},
\author[pisa]{M.~Morganti},
\author[dapnia]{L.~Moscoso},
\author[infn-catania-lns]{M.~Musumeci},
\author[erlan]{C.~Naumann},
\author[erlan]{M.~Naumann-Godo},
\author[cppm]{E.~Nezri},
\author[cppm]{V.~Niess},
\author[nikhef]{G.J.~Nooren},
\author[leeds]{P.~Ogden},
\author[ires]{C.~Olivetto},
\author[dapnia]{N.~Palanque-Delabrouille},
\author[infn-catania-lns]{R.~Papaleo},
\author[cppm]{P.~Payre},
\author[dip-catania]{C.~Petta},
\author[infn-catania-lns]{P.~Piattelli},
\author[ires]{J-P.~Pineau},
\author[dapnia]{J.~Poinsignon},
\author[dip-bologna,iss]{V.~Popa},
\author[cppm]{R.~Potheau},
\author[ires]{T.~Pradier},
\author[ires]{C.~Racca},
\author[infn-catania-lns]{G.~Raia},
\author[dip-catania]{N.~Randazzo},
\author[ific]{D.~Real},
\author[nikhef]{B.A.P.~van~Rens},
\author[cppm]{F.~R\'ethor\'e},
\author[infn-catania-lns]{G.~Riccobene},
\author[toulon]{V. Rigaud},
\author[dip-genova]{M.~Ripani},
\author[ific]{V.~Roca-Blay},
\author[brest]{J-F.~Rolin},
\author[dip-bari]{M.~Romita},
\author[leeds]{H.J.~Rose},
\author[itep]{A.~Rostovtsev},
\author[dip-bari]{M.~Ruppi},
\author[dip-catania]{G.V.~Russo},
\author[dapnia]{Y.~Sacquin},
\author[ific]{F.~Salesa},
\author[erlan]{K.~Salomon},
\author[dapnia]{S.~Saouter},
\author[infn-catania-lns]{P.~Sapienza},
\author[erlan]{R.~Shanidze},
\author[dapnia]{J-P.~Schuller}\footnote{Now at: Istituto di Fisica Nucleare, Sezione di Roma, Piazzale Aldo Moro 2, 00185, Italy},
\author[oxford]{W.~Schuster},
\author[dip-bari]{I.~Sokalski},
\author[dip-bologna]{M.~Spurio},
\author[dapnia]{T.~Stolarczyk},
\author[grphe]{D.~Stubert},
\author[dip-genova]{M.~Taiuti},
\author[sheffield]{L.F.~Thompson},
\author[oxford]{S.~Tilav},
\author[toulon]{P.~Valdy}, 
\author[dip-roma]{V.~Valente},
\author[dapnia]{B.~Vallage},
\author[dapnia]{P.~Vernin},
\author[llv]{J.~Virieux},
\author[nikhef]{G.~de~Vries},
\author[nikhef]{P.~de~Witt~Huberts},
\author[nikhef]{E.~de~Wolf},
\author[itep]{D.~Zaborov},
\author[dapnia]{H.~Zaccone},
\author[itep]{V.~Zakharov},
\author[ific]{J.D.~Zornoza}\footnote{Corresponding author. \\
{\em E-mail address:} zornoza@ific.uv.es \\ 
Now at: University of Wisconsin,
Space Science \& Engineering Center, 222~W~Washington Ave., Madison,
WI~53703, USA},
\author[ific]{J.~Z\'u\~niga}
\end{small}
\end{center}
\address[ific]{IFIC -- Instituto de F\'{\i}sica Corpuscular, Edificios Investigaci\'on de Paterna, CSIC -- Universitat de Val\`encia, Apdo. de Correos 22085, 46071 Valencia, Spain}
\address[grphe]{GRPHE -- Groupe de Recherche en Physique des Hautes Energies, Universit\'e de Haute Alsace, 61 Rue Albert Camus, 68093 Mulhouse Cedex, France}
\address[dip-roma]{Dipartimento di Fisica dell'Universit\`a "La Sapienza" e Sezione INFN, P.le Aldo Moro 2, 00185 Roma, Italy}
\address[lam]{LAM -- Laboratoire d'Astrophysique de Marseille, CNRS/INSU - Universit\'e de Provence Aix-Marseille I, Traverse du Siphon -- Les Trois Lucs, BP 8, 13012 Marseille Cedex 12, France}
\address[dip-genova]{Dipartimento di Fisica dell'Universit\`a e Sezione INFN, Via Dodecaneso 33, 16146 Genova, Italy}
\address[erlan]{University of Erlangen,Friedrich-Alexander Universit\"at Erlangen-N\"urnberg, Physikalisches Institut,Erwin-Rommel-Str. 1, 91058 Erlangen, Germany}
\address[dapnia]{DSM/DAPNIA -- Direction des Sciences de la  Mati\`ere, D\'epartement d'Astrophysique de Physique des Particules de  Physique Nucl\'eaire et de l'Instrumentation Associ\'ee, CEA/Saclay, 91191 Gif-sur-Yvette Cedex, France}
\address[cppm]{CPPM -- Centre de Physique des Particules de Marseille, CNRS/IN2P3 Universit\'e de la M\'editerran\'ee Aix-Marseille II, 163 Avenue de Luminy, Case 907, 13288 Marseille Cedex 9, France} 
\address[oxford]{University of Oxford, Department of Physics, Denys Wilkinson
Building, Keble Road, Oxford OX1 3RH, United Kingdom}
\address[dip-bologna]{Dipartimento di Fisica dell'Universit\`a e Sezione INFN, Viale Berti Pichat 6/2, 40127 Bologna, Italy}
\address[dip-bari]{Dipartimento Interateneo di Fisica e Sezione INFN, Via E. Orabona 4, 70126 Bari, Italy} 
\address[com]{COM -- Centre d'Oc\'eanologie de Marseille, CNRS/INSU Universit\'e de la M\'editerran\'ee Aix-Marseille II, Station Marine d'Endoume-Luminy, Rue de la Batterie des Lions, 13007 Marseille, France} 
\address[nikhef]{NIKHEF, Kruislaan 409, 1009 SJ Amsterdam, The Netherlands}
\address[leeds]{University of Leeds, Department of Physics and Astronomy, Leeds LS2 9JT, United Kingdom}
\address[dip-catania]{Dipartimento di Fisica ed Astronomia dell'Universit\`a e Sezione INFN, Viale Andrea Doria 6, 95125 Catania, Italy}
\address[sheffield]{University of Sheffield, Department of Physics and Astronomy, Hicks Building, Hounsfield Road, Sheffield S3 7RH, United Kingdom}
\address[pisa]{Dipartamento di Fisica dell'Universit\`a e Sezione INFN, Largo
B.~Pontecorvo 3, 56127 Pisa, Italy}
\address[tesre]{IASF/CNR, 40129 Bologna, Italy} 
\address[llv]{UMR GéoScience Azur, Observatoire Océanologique de Villefranche, BP48, Port de la Darse, 06235 Villefranche-sur-Mer Cedex, France}
\address[brest]{IFREMER -- Centre de Brest, BP 70, 29280 Plouzan\'e, France}
\address[toulon]{IFREMER -- Centre de Toulon/La Seyne Sur Mer, Port
  Br\'egaillon, BP507, 83500 La Seyne Sur Mer, France}
\address[ires]{IReS -- Institut de Recherches Subatomiques, CNRS/IN2P3 et Universit\'e Louis Pasteur, BP 28, 67037 Strasbourg Cedex 2, France}
\address[apc]{Universit\'e Paris VII, Laboratoire APC, UFR de Physique,
2 Place Jussieu, 75005 Paris, France}
\address[itep]{ITEP -- Institute for Theoretical and Experimental Physics, B.~Cheremushkinskaya 25, 117259 Moscow, Russia}
\address[infn-catania-lns]{INFN -- Labaratori Nazionali del Sud (LNS), Via S. Sofia 44, 95123 Catania, Italy}
\address[iss]{ISS -- Institute for Space Siences, 77125 Bucharest --
Magurele, Romania}

\begin{frontmatter}
\begin{abstract}
The ANTARES neutrino telescope, to be immersed depth in the
Mediterranean Sea, will consist of a 3 dimensional matrix of 900 
large area photomultiplier tubes housed in pressure resistant glass spheres. 
The selection of the optimal photomultiplier was a critical step for the 
project and required an intensive phase of tests and developments carried out
in close collaboration with the main manufacturers worldwide. This paper
provides an overview of the tests performed by the collaboration and
describes in detail the features of the \pmt\ chosen for ANTARES.

\end{abstract}

\end{frontmatter}

\section{Introduction}

The ANTARES collaboration is constructing a neutrino
telescope~\cite{bib:proposal} to be operated at great depth in the \med. This
telescope is a first step towards a full scale km$^{3}$~detector. To this
end, during the last years many efforts have been put into R\&D studies,
especially concerning one of the most crucial components of the detector, the
photomultiplier tubes (PMTs). The main results of these tests, performed in
close collaboration with PMT manufacturers, are described in this paper.

Section~\ref{sec:principle} describes the detection principle of a deep-sea
neutrino telescope and the layout of ANTARES.  The requirements that the
photo-sensitive detectors should satisfy and other relevant PMT
characteristics are reviewed in section~\ref{sec:specification}.  These
criteria led to the pre-selection of a limited number of PMT candidates. The
results of the main tests performed on these candidates are given in
section~\ref{sec:comparison}.  In the last section, the main features of the
PMT chosen for ANTARES, the Hamamatsu R7081-20, are summarised. These results
are based on the measurements made on a sample of 912 units purchased and
tested in 2001.

\section{Detection principle of a neutrino telescope}
\label{sec:principle}

High energy muon neutrinos can  be detected by observing 
the long-range muons produced by neutrino interactions with the
matter surrounding the detector~\cite{bib:markov}. In a transparent medium,
such as water or ice, the Cherenkov light
induced by these relativistic muons can be detected 
by a three-dimensional array of PMTs.  
The measurement of the arrival time of the Cherenkov light at each PMT
combined with the knowledge of their positions
allows the reconstruction of the muon direction.
The number of photons hitting each PMT gives an estimate of the muon 
energy.

Though simple in principle,  the design of the telescope is 
limited in practice by two essential constraints:
\begin{itemize}
\item  A very large muon flux is naturally  
produced by cosmic ray interactions with the Earth's atmosphere. In order to
reduce this background, the neutrino telescope must be located in the sea (or
ice) at a depth of a few kilometres. Therefore, the PMTs must be housed in a
transparent and pressure-resistant glass sphere. This glass sphere, together
with the equipment inside, is called Optical Module (OM).
\item  Due to the very low flux of high-energy neutrinos from astrophysical
  sources and to the small neutrino interaction cross-section, a large-volume
  detector is necessary.
\end{itemize}

\begin{figure}
\vspace{-1cm}
 \begin{center}
 \mbox{\epsfig{file=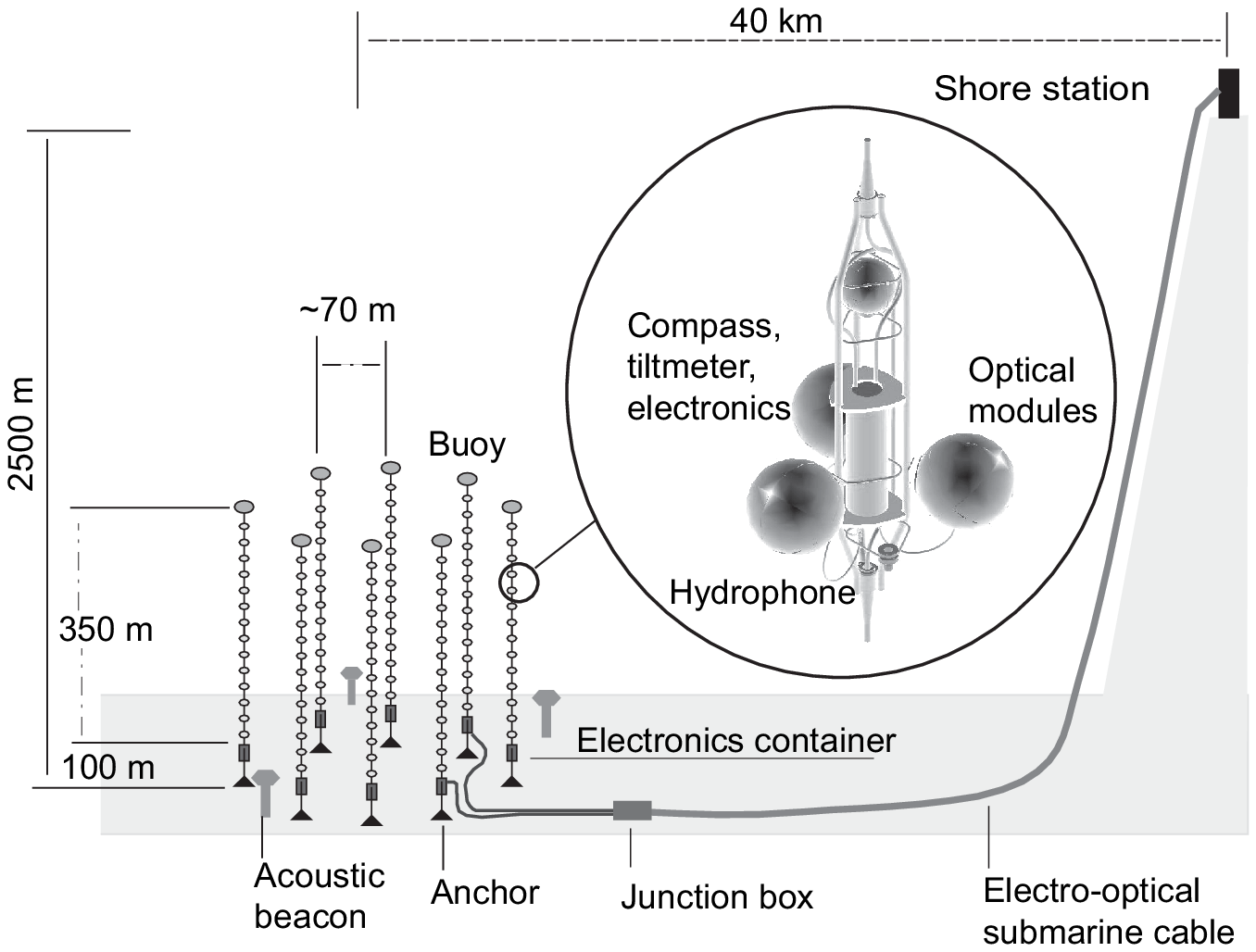,width=0.8\linewidth}}
 \end{center}
 \caption{The ANTARES detector will consist of 12 lines, about 450~m long,
 out of which 350~m are equipped with 75 optical modules grouped in 25
 triplets called storeys.}
 \label{fig:detector}
\end{figure}

Figure~\ref{fig:detector} shows a schematic view of the ANTARES telescope.
It consists of 12 vertical lines, separated by 60-75~m from each other. Each
line is equipped with 75~OMs, in groups of three forming 
storeys separated by 14.5~m.  The whole detector is
immersed at a depth of  2500~m and connected to the shore by a 40~km long
electro-optical cable.

\section{Specifications for the ANTARES photomultiplier tubes}
\label{sec:specification}

  This section describes the specifications required for the photomultiplier
  tubes used in ANTARES. In order to establish a standard set of measurement
  parameters, we require the following conditions during the tests:

\renewcommand{\labelitemi}{-}
\begin{itemize}
  \item an average light level corresponding to well below one photoelectron is used;
  \item the high voltage is fixed to HV$_{nom}$ (see below);
  \item the PMT is shielded against the Earth's magnetic field;
  \item measurements are performed at room temperature ($\sim$20$^\circ$C);
  \item the light source illuminating the photomultiplier should have a
  wavelength of about 450 nm;
  \item the temporal width of the light pulse is lower than 1~ns (FWHM);
  \item light homogeneously illuminates the photocathode.
\end{itemize} 

\renewcommand{\labelitemi}{$\bullet$}
\begin{itemize}

\item{{\bf Dimensions\vspace{0.2cm}}

Figure~\ref{fig:om} shows a sketch of the ANTARES OM \cite{bib:ompaper}. 
It consists of a pressure-resistant glass 
sphere housing the photomultiplier embedded in 
silicon gel to ensure a good optical coupling. 
A high permittivity alloy cage surrounds the tube, shielding it
against the Earth's magnetic field. 
Signal outputs and HV control and monitoring 
are transmitted through  pressure-resistant connectors to 
and from the outside world. The glass sphere, made of two separate halves,
is closed by applying an inner under-pressure of 200-300~mbar. The sphere
dimension (41.7~cm inner diameter) limits the photocathode diameter to
$\sim$38~cm (15'' PMT), and the total PMT length to $\sim$35~cm.}\\

\begin{figure}
\centering
 \mbox{\epsfig{file=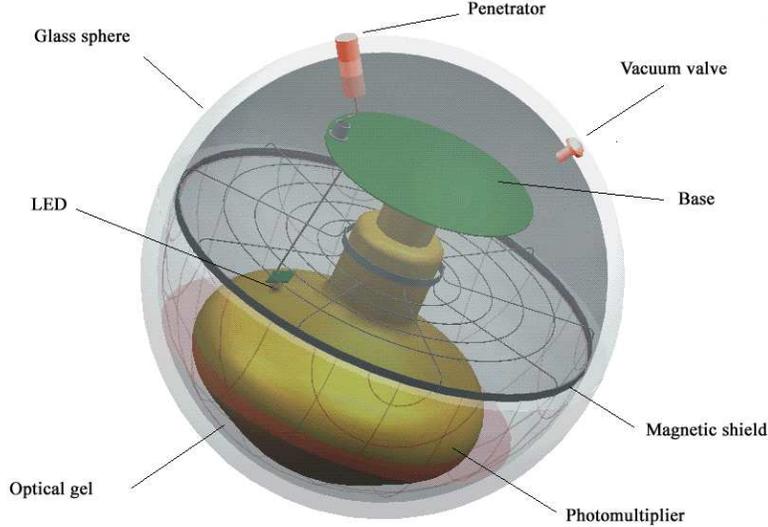,width=0.9\linewidth}}
\caption{Sketch of an ANTARES optical module. A large hemispherical
  photomultiplier (10 inch diameter) is protected by a  pressure-resistant
  glass sphere. The outer diameter of the sphere is 43.2 cm.}
\label{fig:om}
\end{figure}

\item{{\bf Gain\vspace{0.2cm}}

The noise level of the electronic circuit used to digitise the PMT pulse
requires a single photoelectron (SPE) amplitude of the order of 50 mV on a 50
$\Omega$ load~\cite{bib:ars}.  This requires the photomultiplier
gain to be at least $5\times 10^7$.  The HV value corresponding to
this gain will be called nominal voltage, HV$_{nom}$, hereafter.  The
HV$_{nom}$ is not limited in principle, but should remain small enough
($<$2000~V) to ensure long term stability and avoid ageing
problems. Experience shows that these requirements restrict the choice to
tubes equipped with at least 10 amplification stages, linear focusing or
box-line type.}\\

\item{{\bf Peak to valley ratio\vspace{0.2cm}}

  For an absolute calibration of the telescope, it is necessary to be able to
  properly isolate a single photoelectron from the pedestal. This requires a
  peak to valley ratio (P/V) to be at least 2, as computed from the charge
  spectrum (see section~\ref{sec:comparison} for details).}\\

\item{{\bf Transit time spread\vspace{0.2cm}}

  The muon direction is reconstructed from the arrival time of the Cherenkov
  light at each PMT.  The relationship between the required muon angular
  resolution and the time resolution has been estimated using a full computer
  simulation, taking into account the effect of all timing
  uncertainties. Considering that we are aiming at an absolute time
  calibration better than $\sigma_{cal}=1$~ns, the timing uncertainty due to
  the jitter of the PMT response should also remain of the order of 1~ns, in
  order not to degrade significantly the angular resolution. In the
  following, the so-called transit time spread (TTS) will refer to the
  full-width at half maximum (FWHM) of the transit time of the signal in the
  phototube. This TTS should not exceed 3~ns.}\\

\item{{\bf Signal shape characteristics\vspace{0.2cm}}

  In order to be able to separate events which are close in time, the pulse
shape must be stable, non saturated and smooth. Moreover, to improve the time
resolution and the analysis of the PMT signals, the rise-time (measured
between 10\% and 90\% of maximum), fall-time and pulse width must be
shorter than 5~ns, 15~ns and 12~ns, respectively.}\\

\item{{\bf Dark noise\vspace{0.2cm}}

  The dark noise rate, also known as dark count (DC), does not have a big
  impact on ANTARES compared to other experiments, because of the important
  optical background due to natural radioactivity of potassium and
  bioluminescence. Previous measurements~\cite{bib:biolum} show that the
  continuous rate of single photoelectron events is about 60 kHz on a 10''
  photomultiplier\footnote{PMTs also occasionally detect bioluminescence
  bursts which produce sudden and short rate excursions at several MHz.}.

  The DC is measured at a 0.25 photoelectron threshold level 
after the PMT has been kept in the dark for two hours. 
In these conditions, the DC should be less than 10~kHz.}\\

\item{{\bf Out-of-time pulses\vspace{0.2cm}}

 Prepulses and afterpulses have several origins which can be disentangled if
the time interval that separates them from the main pulse is long
enough. Since they are time correlated with the true pulses, prepulses as
well as afterpulses can prevent a correct reconstruction or mimic real
physical events.

We call t$_0$ the mean arrival time of 
the expected  PMT signal associated with the light pulse. The definitions and
limits for the different out-of-time pulses we have considered are given in
table~\ref{tab:afterdef}. \\

\begin{table}
\begin{center}
\begin{tabular}{|l|c|c|c|}\cline{1-4}
 \bf{Type}            &\bf{Time window}     &\bf{Main pulse} &\bf{Limit} \\ \hline \hline
 Prepulses            & [-100 ns, 10 ns]    & Yes/No         & 1\% \\
 Delayed pulses       & [10 ns, 100 ns]     & No             & 5\% \\
 Afterpulses type 1   & [10 ns, 100 ns]     & Yes            & 1\% \\
 Afterpulses type 2   & [100 ns, 16 $\mu$s] & Yes            & 10\% \\ \hline

\end{tabular}
\vspace{0.1cm}
\caption{\small Definitions and limits (ratio of pulse rate in the time
  interval over the main pulse rate) for different out-of-time pulses. The
  time window is defined with respect to $t_0$. The presence of a pulse in
  the main pulse window (t$_0\pm$10~ns) is also shown.}
\label{tab:afterdef}
\end{center}
\end{table}
}

\item{{\bf Photocathode requirements\vspace{0.2cm}}

  Due to the low density of photomultipliers in the ANTARES telescope (about
10$^{-4}$ PMT/m$^3$), the photon yield produced by cosmic events on the OMs
is very small. Therefore, the photocathode must be as large as possible to
minimise the cost of mechanics, within the constraints on the dimensions
discussed previously. The geometrical surface of the photocathode should be
larger than 500~cm$^2$. On the other hand, the quantum efficiency should be
at least 20\% and the collection efficiency at least 80\% over the whole
photocathode surface.}\\

\item{{\bf Mechanical and environmental strength\vspace{0.2cm}}

  The rather peculiar environment of the photomultipliers has to be taken
 into account in defining their specifications. The photomultipliers should
 be able to bear vibrations of up to 1~g and 1~to~55~Hz, produced during
 transport and deployment (boat, deck, etc.). They should also withstand
 temperature variations from -10$^\circ$C to +60$^\circ$C. Once in place, the
 tubes will stay in a rather quiet surrounding at a constant temperature
 ($\sim$13$^\circ$-14$^\circ$C).}\\

\end{itemize}

There are other features which are not considered specifications but are of
interest to characterize the tubes. The energy resolution is a useful
parameter to estimate the fluctuations in charge of the SPE events. A value
below 40\% is considered good enough for this kind of photomultipliers. The
effective area, $A_{eff}$, is defined as the product of the surface of the
photocathode by the collection efficiency. It should be as large as
possible. Finally, the effect of ageing has also to be measured. The
telescope is designed for a lifetime of about 10~years and the total charge
delivered to the anode is expected to be 0.1-0.2~C/day for each PMT. Although
recovery and repair of the elements of the detector is possible, this is a
costly operation. Therefore, the PMTs should have a good stability and a long
lifetime.

\section{Comparison of PMT candidates}
\label{sec:comparison}

A large variety of PMT models have been tested and characterized taking into
account the constraints listed in the previous section.
Tubes considered by ANTARES either already existed on the market or have been
developed recently and specifically for neutrino telescopes. In this case,
their tests were carried out by ANTARES in close collaboration with the
manufacturers.

Among the available models, we chose those closest to the specifications:
the 10''~Hamamatsu~R7081-20, the 10.6''~Photonis~XP1804/D2 and the
13''~Hamamatsu~R8055. Table~\ref{tab:allpms} lists the characteristics of these
tubes. In the following, detailed results of the measurements performed on
these PMTs are presented.

\begin{table}
\begin{center}
\begin{tabular}{|c|c|c|c|}\cline{1-4}
 \bf{Model} &\bf{Photocathode} &\bf{Number of} &\bf{Dynode} \\
            &\bf{diameter}     &\bf{stages}    &\bf{structure} \\ \hline \hline
 Hamamatsu R7081-20~\cite{bib:hama}   & 10''   & 14  & Box line \\
 Photonis XP1804/D2~\cite{bib:photo} & 10.6'' & 11  & Linear focused  \\
 Hamamatsu R8055~\cite{bib:hama}      & 13''   & 10  & Box line  \\ \hline

\end{tabular}
\vspace{0.1cm}
\caption{\small List of the PMT candidates studied by ANTARES.}
\label{tab:allpms}
\end{center}
\end{table}

\subsection{Experimental setup}
\label{sec:setup}
	
  In this section, the experimental setup used at IFIC (Valencia) during the
  selection stage is described. For a description of the setup at CEA/DAPNIA
  (Saclay) see section~\ref{sec:saclay}, where the results obtained on the
  whole sample of R7081-20 are presented.
  
\begin{figure}
 \begin{center}
 \epsfysize=9.0cm
 \epsffile{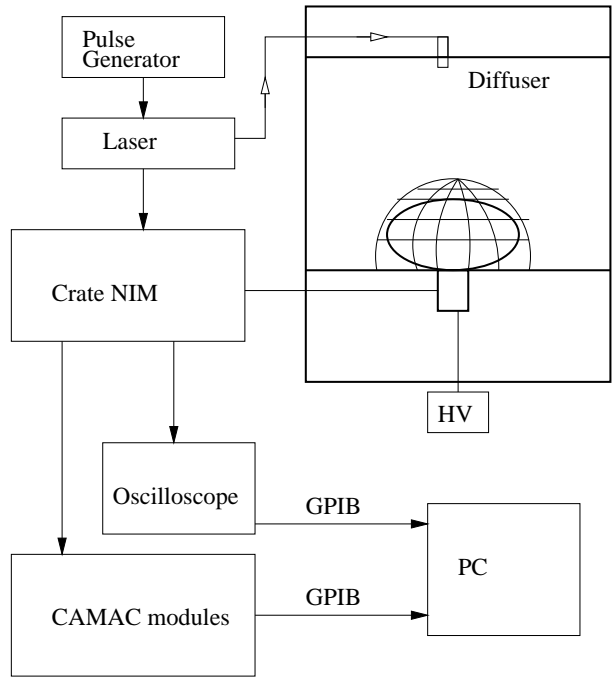}
 \caption{\small Schematic view of the experimental setup. The
 photomultiplier is inside a black box. A $\mu$-metal cage is used in order to
 shield it from the Earth's magnetic field. The laser is
 pulsed by means of a pulse generator and its light is guided via a
 fibre optical cable inside the black box and diffused there to illuminate
 homogeneously the PMT. The acquisition is performed by CAMAC modules or an
 oscilloscope (depending on the measurement) and the data are sent to a PC via GPIB.}
 \label{fig:setup}
 \end{center}
\end{figure}

   As it can be seen in figure~\ref{fig:setup}, the photomultiplier is set in a
   black box (80~cm~$\times$\\~80~cm~$\times$~140~cm) in order to protect it
   from the environmental light. A $\mu$-metal cage shields it against
   the Earth's magnetic field. The HV is supplied by a NIM
   module located  in the crate outside the box. The photomultiplier is
   illuminated by means of a green laser located in another black box and
   triggered by a pulse generator. An optical fibre guides the light from the
   laser into the PMT black box. In order to get uniform illumination of the
   whole photocathode an Oriel~48020 diffuser~\cite{bib:oriel} is used at the
   end of the optical fibre at a distance of about 80~cm from the PMT
   head. Depending on the measurement being performed, the output signal from
   the photomultiplier is sent either to NIM and CAMAC modules or to a
   digital oscilloscope (Tektronix TDS640A~\cite{bib:scope}).
   Both the CAMAC modules and oscilloscope are read out through a GPIB 
   link by a LabView program running  on a PC.

   The green laser used as light source to illuminate the photomultiplier is
   a Nd-YAG device\footnote{ The laser model is NG-10120-120 from
   Nanolase~\cite{bib:nanolase}}, i.e. a synthetic (Y$_{3}$Al$_{5}$O$_{12}$)
   crystal doped by Nd in concentrations of 10$^{20}$ atoms/cm$^3$. It is
   pulsed by the Q-switching mechanism, providing very fast pulses
   ($\sigma\leq$0.5~ns). The emitted energy per pulse is
   $\sim$1.2~$\mu$J. The output stability is better than 1\% over several
   hours and the energy emission changes less than 1\% after switching on and
   off. The emission wavelength is 532~nm. An internal photodiode located in
   its head provides a trigger signal with an error smaller than 50~ps.

\subsection{PMT comparison}
\label{comparison}

The following quantities were measured for each of the 3 PMT candidates:
gain, amplitude, peak to valley ratio, energy resolution, transit time
spread, dark noise rate and out-of-time pulse rate. Much of this information
(gain, peak/valley, energy resolution) is obtained from the SPE charge
spectrum. This distribution can be described by

   \begin{equation}
     \label{eq:espec}
     S(q)=G_{p}(q;\mu_{p},\sigma_{p}) + F(q;\omega,\alpha,\beta) + 
\sum_{n=1}^{\infty} P(n;\mu) \times G_n (q; \mu_{n}, \sigma_{n}), 
   \end{equation}

\noindent where $G_{p}$ and $G_{n}$ are the Gaussian distributions
corresponding to the pedestal and to $n$ photoelectrons respectively and $F$
is the valley function. The valley is approximated by the sum of an
exponential and a Gaussian. Three parameters are needed to describe this
phenomenon: the fraction of badly amplified photoelectrons $\omega$, the
fraction of valley events contributing to the exponential (and not to the
Gaussian) $\beta$ and the ``slope'' of the exponential $\alpha$. $P$ is the
Poisson probability of producing $n$ photoelectrons at the first dynode if
the product of the number of photoelectrons emitted at each laser shot, the
quantum efficiency and the collection efficiency is $\mu$. If $\mu_{1}$ and
$\sigma_{1}$ are the parameters of the Gaussian for a single photoelectron
reaching the first dynode, the corresponding parameters for two or more ($n$)
photoelectrons are $\mu_{n}=n \cdot \mu_{1}$ and $\sigma_{n}^{2}=n \cdot
\sigma_{1}^{2}$.

\begin{figure}
 \begin{center}
 \epsfysize=7.0cm
 \epsffile{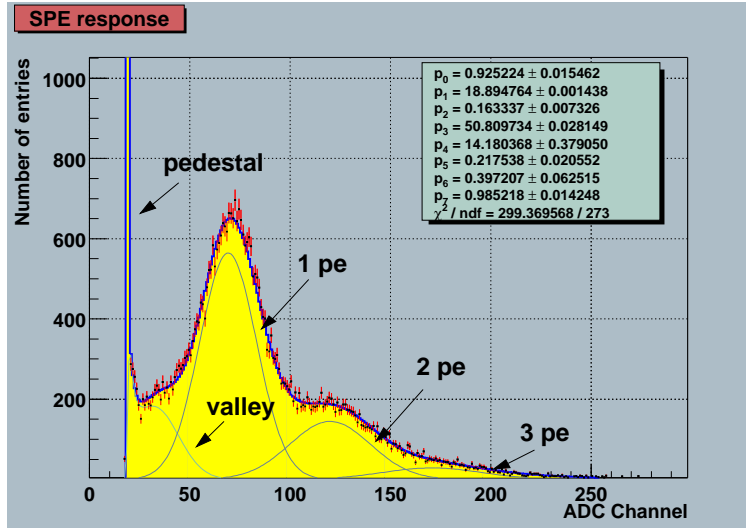}\\
 \caption{\small Example of charge spectrum. Each of the contributions
 described in the text is shown separately. Each channel of the ADC
 corresponds to 0.25~pC.}
 \label{fig:ej_spe}
 \end{center}
\end{figure}

Therefore, there are 8 free parameters ($\mu, \mu_{p}, \sigma_{p}, \mu_{1},
\sigma_{1}, \omega, \alpha, \beta$) that can be obtained fitting the measured
charge spectrum to equation~\ref{eq:espec}. A typical example of a charge
spectrum is shown in figure~\ref{fig:ej_spe}.

\subsubsection{Gain}
\label{gain}

\begin{figure}
 \begin{center}
 \epsfxsize=10cm
 \epsffile{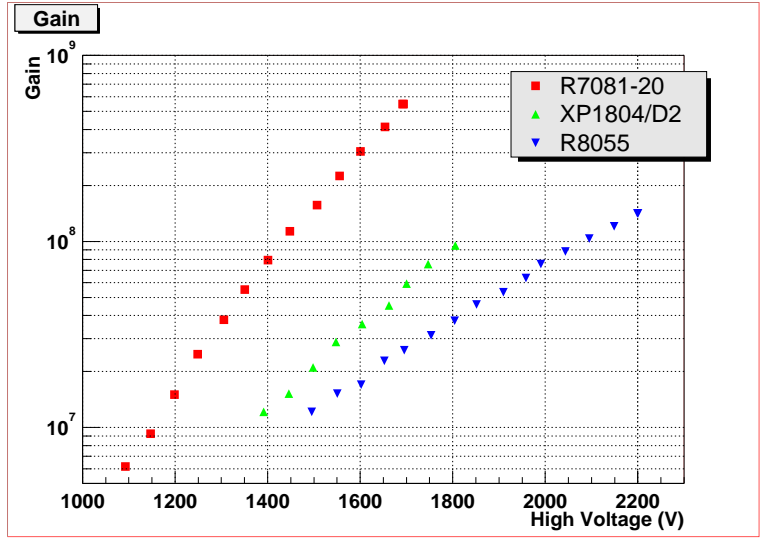}
 \caption{\small Gain dependence on HV for the three candidate photomultipliers.}
 \label{fig:gain-volt}
 \end{center}
\end{figure}

The gain is obtained from the position of the SPE
peak in the charge spectrum. Figure~\ref{fig:gain-volt} shows the gain
dependence on HV for the three candidates. From these plots, the nominal
voltage can be determined: 1340~V for the R7081-20, 1680~V for the XP1804/D2
and 1900~V for the R8055.

The highest gains are reached by the Hamamatsu R7081-20 model, as expected
since this tube has 14 dynodes. It was designed to attain gains in excess of
$10^9$, required by some experiments to transfer the analogical PMT signal
along km-long electric cables (AMANDA~\cite{amanda}).  In the case of the
Photonis XP1804/D2, which has only 11~dynodes, the nominal voltage is
naturally higher and the increase in gain with HV smaller than for the
R7081-20 model.  Finally, the Hamamatsu R8055 needs an even higher voltage to
reach the nominal gain. This is not surprising since the HV division for this
tube was optimized for TTS and not for gain. As we will see later, all 3 PMTs
are able to reach a gain of $10^{8}$ without significant deterioration of
their properties.

\subsubsection{Amplitude}
\label{amplitude}

The experimental setup used to obtain the SPE amplitude distribution is
similar to that used for the charge spectrum measurement. In this
case, the acquisition is performed by a digital oscilloscope. 

\begin{figure}
 \begin{center}
 \epsfxsize=10.0cm
 \epsffile{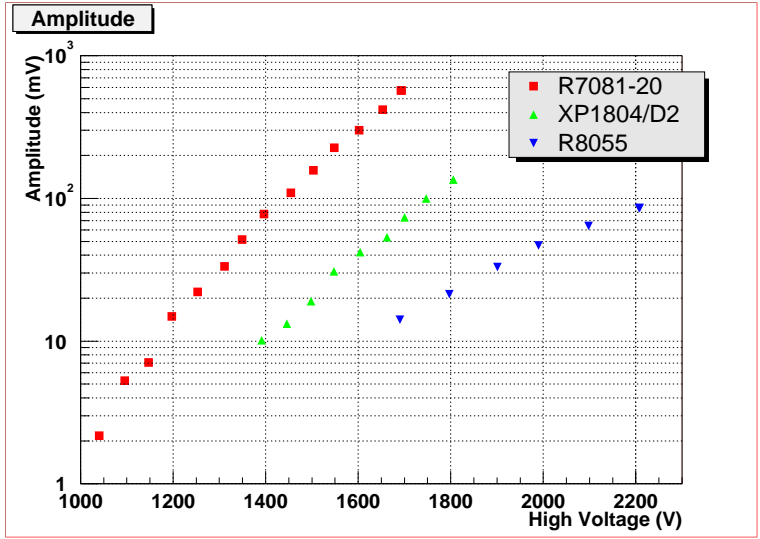}
 \caption{\small Amplitude dependence on HV for the three candidate photomultipliers.}
 \label{fig:ampli-volt}
 \end{center}
\end{figure}

The fitting procedure is simpler than in the previous case. The SPE peak  is
fitted to a Gaussian, whose mean is the amplitude. Figure~\ref{fig:ampli-volt}
shows that the dependence of the amplitude on HV is very similar to that of the charge. 
This is to be expected since for a given PMT, amplitude and charge are nearly
proportional and related to the pulse shape which in turn is linked to 
the dynode structure of the photomultiplier. It is important to
note that in the ANTARES experiment, the signal amplitude will be
used in the trigger (see~\cite{bib:ars} for a more detailed description).

The typical SPE amplitudes between anode and ground on 50~$\Omega$ at the
nominal voltage are 46,~60~and~34~mV respectively for the R7081-20, the
XP1804/D2 and the R8055.

\subsubsection{Peak to valley ratio}
\label{pv}

\begin{figure}
 \begin{center}
 \epsfxsize=10.0cm
 \epsffile{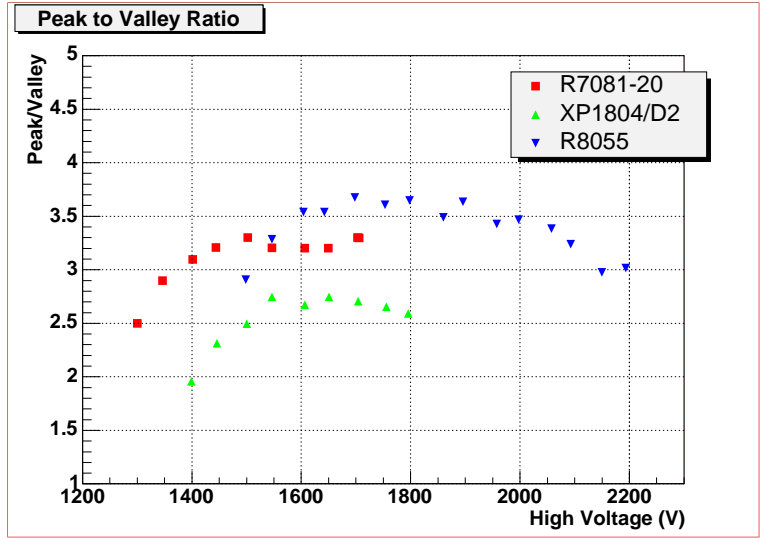}
 \caption{\small Peak to valley ratio dependence on HV for the three candidate PMTs.}
 \label{fig:pvs}
 \end{center}
\end{figure}

  The ratio between the height of the SPE peak and the valley, measured from
  the SPE charge spectrum, is important since it can be used to estimate the
  capability of the photomultiplier to discriminate the number of
  photoelectrons in an event. As it was seen in
  section~\ref{sec:specification}, the ANTARES specifications require a P/V
  larger than~2. Figure~\ref{fig:pvs} shows the results of the measurements
  for this quantity.

The peak to valley ratio of the Hamamatsu R7081-20 model at the nominal
voltage is around 2.8. For higher HV, it is almost flat in the region of
interest.
Concerning the Photonis photomultipliers, it can be seen
that the P/V ratio is around 2.7. It grows with HV at low voltages 
and then levels off. The peak/valley
ratio of all the Hamamatsu R8055 tubes is very good (P/V=3.7) 
and stable near the nominal voltage. For very low or very high 
voltages, the peak/valley ratio decreases.

 \subsubsection{Energy resolution}
\label{ser}
\begin{figure}
 \begin{center}
 \epsfxsize=10.0cm
 \epsffile{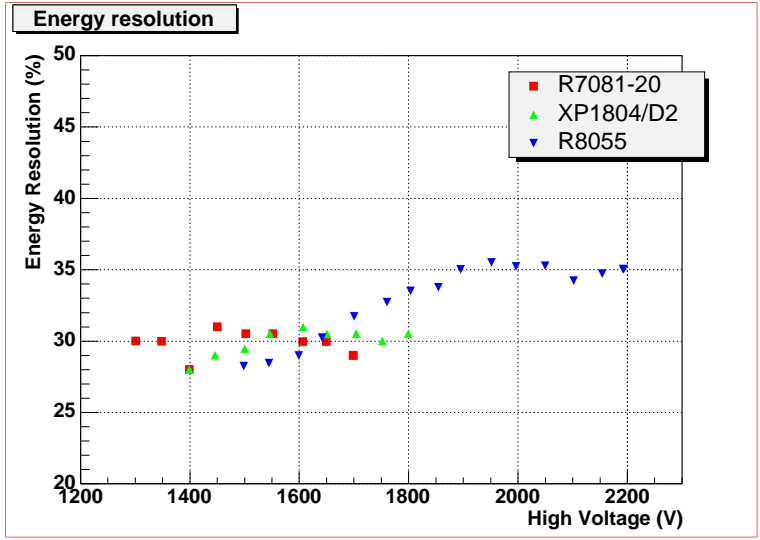}
 \caption{\small Energy resolution dependence on HV for the three candidate PMTs.}
 \label{fig:eners}
 \end{center}
\end{figure}

Energy resolution is calculated as the ratio of $\sigma_{1}$ to $\mu_{1}$. The
results are shown in figure~\ref{fig:eners}.
 
The 10'' Hamamatsu model exhibits an almost flat dependence on HV. The
behaviour of the XP1804/D2 is also smooth. There is a slow increase of the
energy resolution with HV, which becomes stable near the nominal voltage. At
this point, energy resolution is in general over 30\%. In the curve
corresponding to the R8055, two regions can be observed: the energy
resolution undergoes a relatively fast increase at low voltages and then
stabilises. Table~\ref{tab:nom_volt} shows the energy resolution at the
nominal voltage.

\subsubsection{Transit time spread}
\label{stts}

As already pointed out, the transit time spread is a key parameter and has to
be measured with high accuracy. The setup for this measurement is as
follows. The laser illuminates the photomultiplier, whose output is
discriminated in order to provide the STOP signal to the TDC. The START
signal is given by the internal photodiode of the laser. The results
presented in this section were obtained setting the discrimination threshold
at 0.5~pe. The total FWHM consists of three components:
$\Gamma^{2}=\Gamma_{P}^{2}+\Gamma_{L}^{2}+\Gamma_{e}^{2}$, where $\Gamma_{P}$
is the intrinsic TTS of the photomultiplier that we aim to measure,
$\Gamma_{L}$ is the FWHM of the laser pulse and $\Gamma_{e}$ is the time
jitter contribution from the electronics. In order to subtract the two last
terms, a very fast photomultiplier was used. Hamamatsu indicates for this
photomultiplier (model R5496) a value of FWHM~$=0.23\pm0.01$ ns. Illuminating
the fast photomultiplier using the same experimental setup it is possible to
measure the total width of the distribution $\Gamma$ and calculate the effect
of the source and the electronics.  Our measurement shows that they
contribute as much as $0.65\pm0.05$~ns to the total FWHM.

\begin{figure}
 \begin{center}
 \epsfxsize=10.0cm
 \epsffile{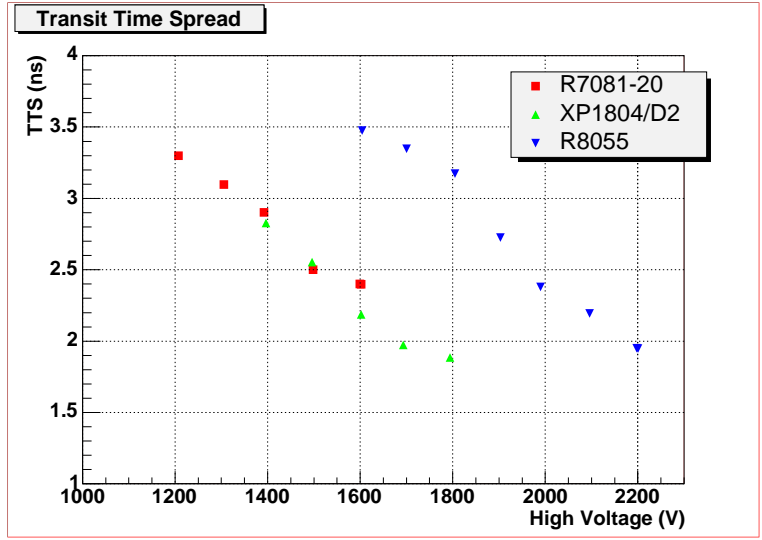}
 \caption{\small TTS dependence on HV for the three candidate photomultipliers.}
 \label{fig:tts}
 \end{center}
\end{figure}

The dependence of the TTS on HV is shown in
figure~\ref{fig:tts}. As  expected, the TTS decreases for higher voltages.
This behaviour is nearly the same for all three candidates, with a decrease of
about 1~ns for a 300-400~V change around the nominal voltage.

As in the case of the charge spectrum, the errors given by the fit are much
smaller than those due to fluctuations in the experimental conditions, so
that an estimate of the latter is obtained by repeating the measurements. The
best values of TTS are obtained for the Photonis PMT. However, all the models
provide results within the chosen specifications.

\subsubsection{Dark noise}
\label{sdark}

To measure the dark current, the output from the photomultiplier is sent to a
Linear Fan In / Fan Out module that sends identical signals to amplitude
discriminators. The number of pulses above each threshold is counted. Each
discriminator uses a different threshold. For the sake of comparison among
the tubes, this threshold is expressed in number of photoelectrons
in amplitude.

\begin{figure}
 \begin{center}
 \epsfxsize=10.0cm
 \epsffile{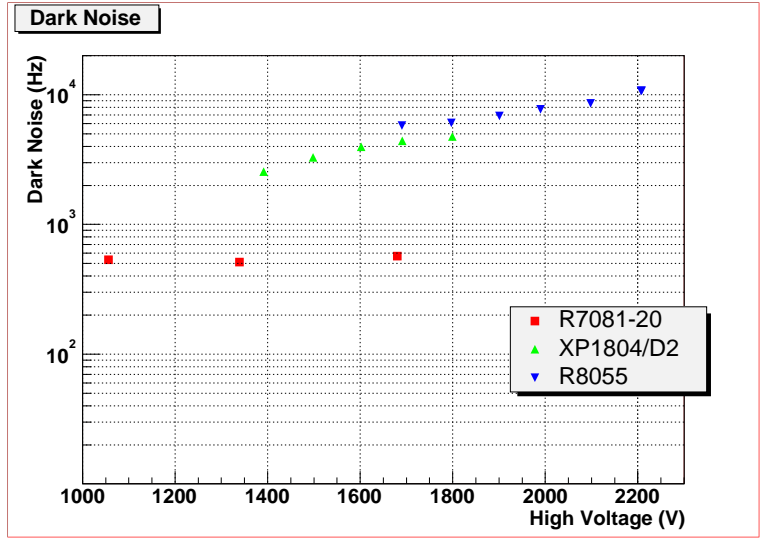}
 \caption{\small Dark noise rate dependence on HV. A threshold of
 half a photoelectron has been  used.}
 \label{fig:darkrate_hv}
 \end{center}
\end{figure}

The dependence of the dark noise on HV has been measured using this setup.
The results are plotted in figure~\ref{fig:darkrate_hv}, where the dark noise
rate for a threshold of half a photoelectron is shown.

Whereas the dark noise dependence on the HV for the 10'' Hamamatsu PMT is
almost flat, it increases smoothly for the other two tubes. This increase is
faster at low voltages for the 10.6'' Photonis PMTs, while the opposite
occurs for the 13'' Hamamatsu tubes, where the increase is faster at high
voltages.

\subsubsection{Out-of-time pulses}

 As it has been mentioned,
  it is important to control out-of-time pulses since it may cause
  misreconstruction problems. Several kinds
  of events have been analyzed.

  The experimental setup used to measure the out-of-time pulse rate is
  similar to that used to measure the SPE spectrum. The photomultiplier is
  illuminated by a low level of light ($\sim$0.1~pe per pulse) and its output
  is sent to the oscilloscope, which is triggered by the laser internal
  photodiode. The oscilloscope digitizes the output from the PMT and sends it
  to the PC, where a LabView program finds the peaks in the waveform. The
  time scale used in the analysis of prepulses, delayed pulses and afterpulses-1
  is 0.2~ns per channel which allows to cover $\pm$200~ns around the main
  pulse. On the other hand, the scale to measure afterpulses-2 is 10~ns per
  channel, i.e. it covers the region from $-4$~to~16~$\mu$s with respect to
  the main pulse window. In this case, the signal from the photomultiplier is
  too narrow to be detected by the oscilloscope. A discriminator
  is used, which implies, however, a loss of the amplitude information.

  The R7081-20 is within the specifications. The
  XP1804/D2 shows a high rate of afterpulses-1 and -2. Some of the
  rates for the R8055 are also high (see table~\ref{tab:nom_volt}).

\subsubsection{Photocathode effective area}

As mentioned in section~\ref{sec:specification}, the effective area is the surface
of the photocathode multiplied by the collection efficiency, which is often
dependent on the hit position. This dependence is estimated by scanning the
PMT with a collimated LED, assuming the collection efficiency to be near
unity at the very centre of the photocathode.

As expected, the best value of the effective area
corresponds to the larger PMT, i.e. the Hamamatsu R8055. 
The smallest effective area is that of
the Photonis XP1804/D2, although being a bit larger in diameter than the
Hamamatsu R7081-20.

\subsection{Conclusions on comparison}

Tables~\ref{tab:nom_volt}~and~\ref{tab:others} summarize the value of the
measured parameters. All the candidates present good performance in
general. All PMTs have good values for gain (the nominal voltage is lower
than 2000~V and the gain reaches 10$^8$). The peak to valley ratio for all
the models is well in agreement with our requirements for all the models,
being always above 2. The values of the energy resolution are similar, and
all within the specifications. Concerning the transit time spread, the
XP1804/D2 shows the lower value of TTS, the values for R7081-20 and R8055
being also acceptable. The best value of dark noise is achieved by the
R7081-20 (better than expected given its smaller photocathode). The other two
models are also within the specifications.  Both XP1804/D2 and R8055
present high afterpulse rates and the latter has also a high prepulse
rate. The PMT with a best ratio of the photocathode effective area to the
photocathode surface is the Hamamatsu R7081-20. Finally, the Hamamatsu
R7081-20 model was chosen for the ANTARES detector.

\begin{table}
\begin{center}
\begin{tabular}{|c|c|c|c|c|}\cline{1-5}
 \bf{Model}           & \bf{ Specification}  &\bf{ R7081-20}     &\bf{XP1804/D2}  & \bf{R8055}  \\
\hline \hline
Nominal Voltage (V)    &  $\leq$2000   &  1340             &   1680         &   1900      \\
Peak/Valley            &  $\geq2$      &  2.8              &    2.7         &    3.7      \\
TTS (ns)               &  $\leq$3      &  3.0              &    2.0         &    2.6      \\
Dark Noise (Hz)        &  $\leq$10000    &  910     &    9200        &   8100      \\
Prepulses (\%)         &  $\leq$1      &  0.01             &    0.01        &    2.2*      \\
Delayed pulses (\%)    &  $\leq$5      &  3.6              &    4.7         &    5.0      \\
Afterpulses-1 (\%)     &  $\leq$1      &  1.0              &    3.2*        &    2.4*      \\
Afterpulses-2 (\%)     &  $\leq$10     &  3.8              &    18*         &    19*       \\ \hline
\end{tabular}
\vspace{0.1cm}
\caption{\small Comparison of the specifications and measured values
  (at a gain of $5 \times 10^7$) of the 
parameters of the three candidate photomultipliers.
TTS and dark noise are measured with a threshold of 0.25 pe. The definitions
of the out-of-time pulses are given in the text. The symbol * indicates that
 the value is outside the specifications.}
\label{tab:nom_volt}
\end{center}
\end{table}

\begin{table}
\begin{center}
\begin{tabular}{|c|c|c|c|}\cline{1-4}
 \bf{Model}             &  \bf{ R7081-20}     &\bf{XP1804/D2}  & \bf{R8055}  \\
\hline \hline 
Amplitude (mV)          &     46               &    60          &    34       \\
Energy Resolution (\%)  &     30               &    31          &    35       \\
Effective Area (cm$^2$) &     450              &    410         &   830  \\ \hline
\end{tabular}
\vspace{0.1cm}
\caption{\small Other features of the studied tubes, not considered as specificaions.}
\label{tab:others}
\end{center}
\end{table}

\section{Hamamatsu R7081-20}
\label{sec:R7081}

The results presented in this section correspond to the final sample of 912
PMTs of the Hamamatsu model R7081-20 which have been characterized in
the ANTARES test bench at CEA/DAPNIA, Saclay.

\label{sec:saclay}

Given the high number of PMTs involved, the experimental test setup has to
allow all measurements to be performed at a fast rate. The test bench
consists of a light tight plastic cylinder 1.5~m long and 0.46~m in
diameter. The cylinder was painted in black on the inside to avoid
reflections and covered by a $\mu$-metal foil externally. Light from a pulsed
blue LED is fed through an optical fibre into one extremity of the tube. A
Lambertian diffuser (Oriel 48010~\cite{bib:oriel}) located just after the LED
and about 1.1~m away from the PMT is used to ensure uniform illumination of
the whole photocathode. The PMT itself is shielded by a $\mu$-metal cage that
reduces the Earth's magnetic field by a factor 3 to 4. Instead of the PMT
base provided by Hamamatsu, a new base (a modified version of the PHQ5912
base developed by ISEG~\cite{bib:iseg}) was used. The HV and the distribution
for the dynodes is generated by the base itself. This change also allowed
some improvement in the PMT response, in particular in the TTS. Due to this
new design, the results given in this section for the R7081-20 model slightly
differ from those given in section~\ref{sec:comparison}. A summary of the
results of the parameters studied for this sample of PMTs is shown in
Table~\ref{tab:mean-rms}.

\begin{table}
\begin{center}
\begin{tabular}{|c|c|c|}\cline{1-3}
 \bf{Parameter}         & \bf{ Mean}    & \bf{ RMS} \\ \hline \hline
Nominal Voltage (V)    &  1804         &  57       \\
Amplitude (mV)         &  42.7         &  2.3      \\
Peak/Valley            &  2.79         &  0.46     \\
TTS (ns)               &  2.79         &  0.15     \\
Raise Time (ns)        &  4.30         &  0.10     \\
Fall Time (ns)         &  12.57        &  0.74     \\
Time width (ns)        &  7.48         &  0.35     \\
Dark Noise (Hz)        &  1880         &  1078     \\
Prepulses (\%)         &  0.010        &  0.014    \\
Delayed pulses (\%)    &  3.71         &  0.24     \\
Afterpulses-1 (\%)     &  1.50         &  0.27     \\
Afterpulses-2 (\%)     &  1.23         &  0.44     \\ \hline
\end{tabular}
\vspace{0.1cm}
\caption{\small Mean and RMS of the distributions of several parameters for
the 912 Hamamatsu R7081-20 PMTs.}
\label{tab:mean-rms}
\end{center}
\end{table}

Finally, in order to study the ageing of the selected model, three tubes
were placed in a black box and excited by three independent blue LEDs located
a few centimeters above the top of the PMTs. LEDs were tuned to reproduce the
optical background environment ($^{40}$K and bioluminescence) expected in the
ANTARES site~\cite{bib:biolum}. In a first run, the LEDs were programmed to
emit a continuous rate of 100~kHz and bursts of 1~MHz lasting 0.1 second
every second. The total integrated charge in the PMTs amounts to more than
100~C, which corresponds to two years in the sea at a constant gain of
10$^{8}$. Subsequently, the LED pulse rate was increased by a factor of 5,
thus allowing a total charge of 500~C, corresponding to 10~years in the sea,
to be collected by the PMTs. The most significant variation of the PMT
properties concerns the gain. Over the first 100 days, all three PMTs showed
an increase of 50-70\% in gain, before they level off. This phenomenon
corresponds to the runnning-in phase, rather than ageing. All other important
energy and timing properties of the PMTs (P/V, energy resolution and TTS)
were stable over the test period. The dark current remained also stable with
time. However, afterpulse-2 rate showed an increase (around 30\%) during the
test, possibly caused by the fact that these pulses are usually due to gas of
ionized atoms, whose number increases with time.

\section{Conclusion}
The ANTARES collaboration has carried out a study of large cathode photomultipliers for a submarine neutrino
telescope. Several candidates have been considered. In particular, intensive
tests on the  R7081-20 and R8055 models from Hamamatsu and the XP1804/D2 from
Photonis have been performed. The measurements to compare them show that they
have in general appropriate performance. 
Finally, the R7081-20 model was chosen for the contruction of
the ANTARES neutrino telescope and the full quantity of 912 PMTs
supplied were tested. The measurements on this batch show that the mean
values are within the specifications (only the average afterpulse-1 rate is
slightly over the limit). The tubes which are not within the specifications
are few and close to the limits. Nevertheless they were replaced.



\end{document}